\newcommand\sps{\space\space\space\space}
  \def\selectedoptions{final}

\newcommand{\pom}{I\hspace{-0.1em}P}
\newcommand{\xpom}{\mbox{$x_{_{\pom}}$}}
\newcommand {\zpom} {\mbox{$z_{_{\pom}}$}}

\documentclass[
   \selectedoptions
  ]
  {aipproc}

\layoutstyle{6x9}

\SetInternalRegister\hbadness{8000} 

\newcommand\doingARLO[2][]{%
  \ifx\mmref\undefined #1\else #2\fi
}

\begin{document}

\title 
      [ZEUS Results]
      {ZEUS Results}


\author{Elisabetta Gallo}{
  address={INFN Firenze, Italy},
  altaddress={(On behalf of the ZEUS Collaboration)}
 }

\copyrightyear  {2005}

\begin{abstract}
Several results from the ZEUS Collaboration were presented at this Workshop.
The highlights are presented in this summary, and include results from
NLO QCD fits and determination of $\alpha_{\mathrm S}$, from forward jets 
and diffractive final states, from pentaquarks and searches and 
from heavy flavour production. Also the first
results from the analysis of the HERA II $e^+p/e^-p$ data are shown.  
\end{abstract}

\date{\today}

\maketitle

\section{Introduction}

These proceedings report on the highlights of recent results from the ZEUS 
Collaboration. The published structure functions data have been used
to determine the proton parton distribution functions in NLO QCD fits.
The fit includes also jet data to constrain the gluon in the middle-$x$ region.
The strong coupling constant $\alpha_{\mathrm S}$ is determined 
in various measurements. The diffractive parton distribution functions
are also determined from inclusive data and used to obtain predictions
for diffractive final states. 
ZEUS has performed searches for exotic final states like pentaquarks, involving
strange or charm quarks. The HERA experiments are still competitive in
the search for SUSY with $R$-parity-violating interactions and the recent
search for stop is described. 
Several results on heavy flavour production were
presented in the parallel sections, here the results on beauty cross sections
at HERA I and $D^\ast$ production at HERA II are reported. Finally the
first results on the charged current polarized cross sections are shown.     

\normalfont

\section{Structure functions and jets}

\subsection{NLO QCD fits}

\begin{figure}
{\includegraphics[height=8.5cm]{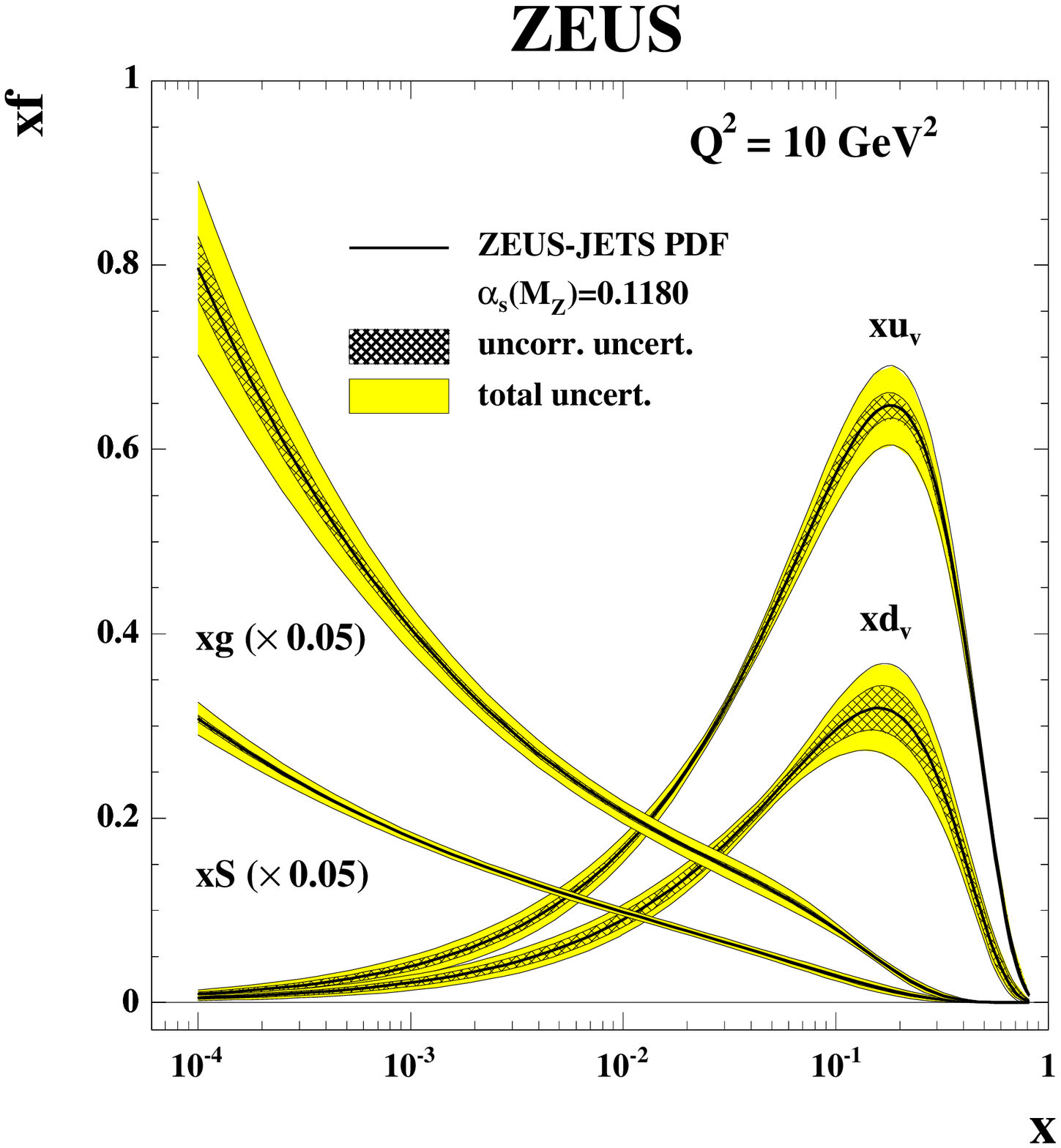}}
{\includegraphics[height=8.5cm]{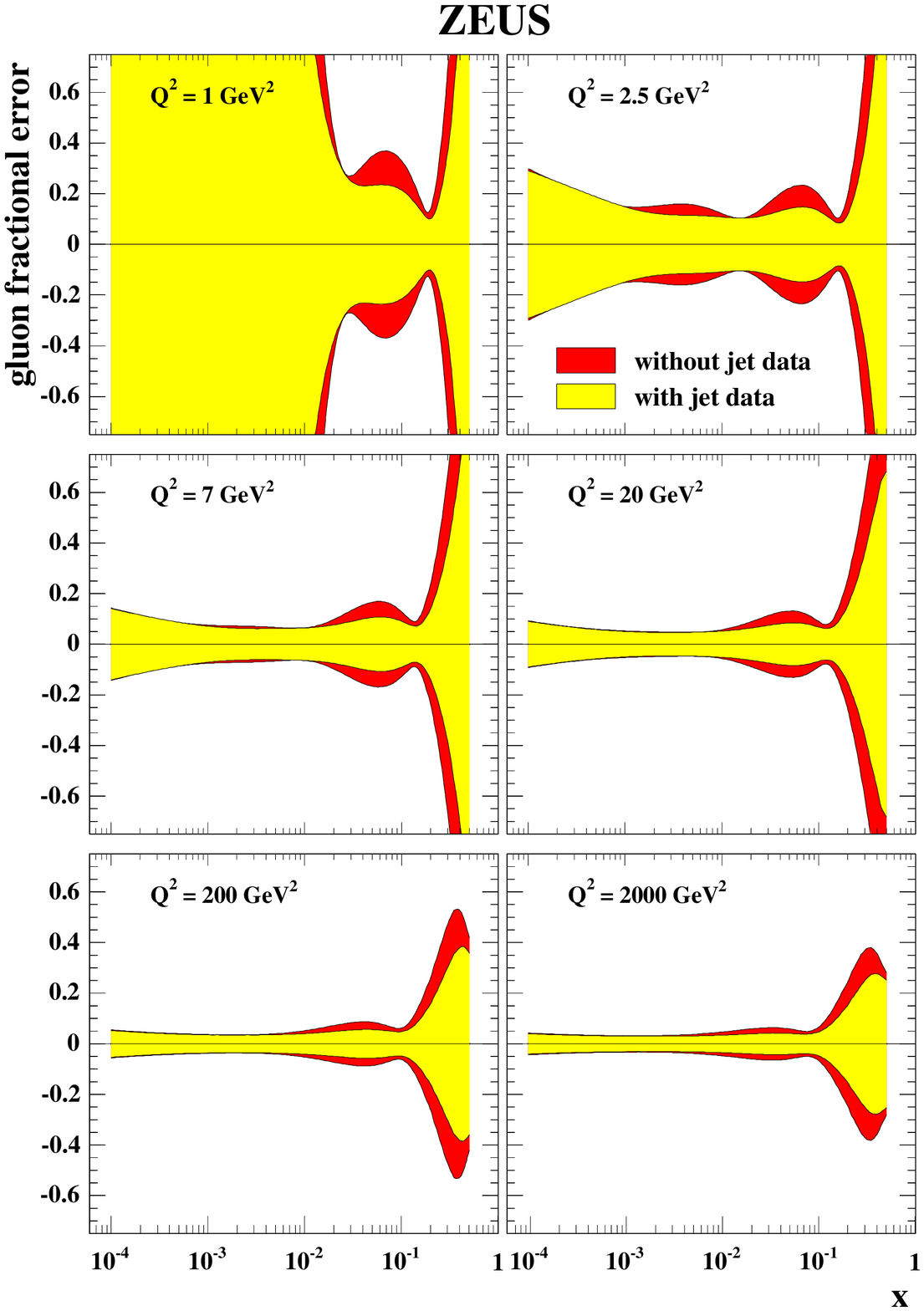}}
  \caption{On the left: parton distribution functions for the $u$-valence, $d$-valence, sea and gluon density obtained from the ZEUS-JETS fit, for a $Q^2$ value
of $10~\mathrm{GeV}^2$. On the right side: uncertainty on the gluon density for different
values of $Q^2$ and for the two cases in which the jets are included or not in the fit.}
  \label{gluon}
\end{figure}

The ZEUS Collaboration has completed the publication of the structure
function data from the HERA I (94-00) running. These cross sections have been
used in a NLO QCD fit, based on the DGLAP evolution, to determine the
proton parton distributions functions (PDFs), using the ZEUS data only
\cite{zeusjets}. 
Uncertainties from heavy-target corrections, present in global
analyses which include also fixed-target data, are therefore avoided.
While the low $Q^2$ neutral current (NC)
data determine the low-$x$ sea and gluon distributions, the high $Q^2$ NC and charged
(CC) cross sections constrain the valence distributions.
In addition, published jet cross sections from the 96-97 data were used to constrain
the gluon density in the mid-to-high-$x$ region ($x \simeq 0.01-0.5$). The predictions
for the two jet cross sections used (DIS jets in the Breit frame 
and jets in direct photoproduction) 
were calculated to NLO and  used in the fit in a rigourous way. The 
resulting PDFs in this fit (see figure~\ref{gluon}), called ZEUS-JETS fit, give a very good description of
both inclusive and jet cross sections, in agreement with QCD factorization.
The gain in using the jets is shown in the right part of figure~\ref{gluon}, where the yellow (light) band
shows the total experiment uncertainty on the gluon density in the fit including the jets, compared to the red
(dark) band which is the error for the fit without including
the jets. As an example, in the bin at $Q^2=7~\mathrm{GeV}^2$ and $x\simeq 0.06$, the uncertainty
is reduced from $17\%$ to $10\%$ using the jets. A similar decrease of approximately a factor two is visible in the whole $Q^2$ range in the mid-to-high-$x$ region.

In the inclusive cross sections, there is a strong correlation between $\alpha_{\mathrm S}$ and the gluon density. As the jets data depend on $\alpha_{\mathrm S}$ and on $x g(x)$ in a different way, the addition of the jets data to the fits permits also a
more precise determination of the strong coupling constant, compared to previous fits, when $\alpha_{\mathrm S}$ is 
left as a free parameter. The value resulting from this fit, called 
ZEUS-JETS-$\alpha_{\mathrm S}$, is
  
\begin{equation}
\alpha_S(M_Z)= 0.1183 \pm 0.0028 (\mathrm{exp.}) \pm 0.0008(\mathrm{model}) 
                       \pm 0.005 (\mathrm{scale})
\end{equation}
which is in very good agreement with the world average of $0.1182 \pm 0.0027$.

\subsection{Summary of $\alpha_{\mathrm S}$ results}

\begin{figure}
{\includegraphics[height=10cm]{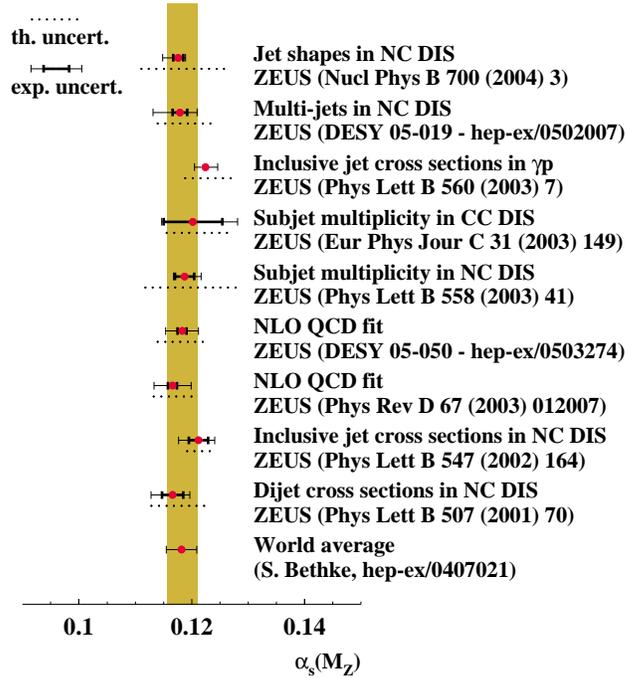}}
  \caption{Compilation of ZEUS results on the determination of
$\alpha_{\mathrm S}$. The band shows the world average as calculated in
the reference cited in the figure.}
 \label{alphas}
\end{figure}

Figure~\ref{alphas} shows a compilation of results on the determination of $\alpha_{\mathrm S}$ at
ZEUS (see also~\cite{claudia}). The most precise result, from the experimental point of view, comes from
the inclusive jets in photoproduction. Although each of the measurements has a precise experimental
determination (typically $3\%$),
 competitive with the world average, the  error is dominated by the theoretical uncertainties (typically $4\%$), which are shown by the dashed lines. NNLO (next-to-NLO) calculations for jet based variables and for PDFs are  therefore needed.


\subsection{Forward jets at low-$x$}

\begin{figure}
{\includegraphics[height=9cm]{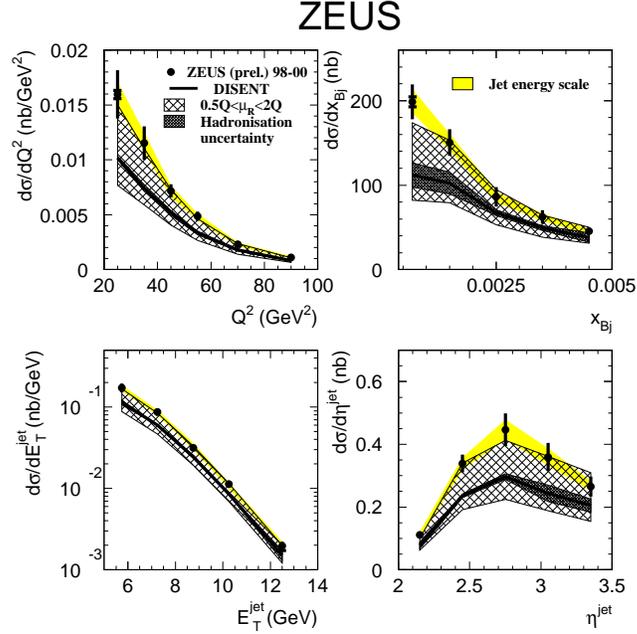}}
  \caption{Measured differential cross sections compared to the NLO QCD calculations for forward jet production at low $x$.}
  \label{fjets}
\end{figure}

The DGLAP evolution scheme describes the HERA structure function data  down to low-$x$. In order to
disentangle the effect of different evolution schemes, like BFKL or CCFM, it has been suggested long ago by Mueller~\cite{mueller} to look at jets in the forward (proton) region, where the differences between the different parton
evolution schemes should be more prominent. For such an analysis, 
the jets were reconstructed with the $k_T$ cluster algorithm in the
longitudinally invariant inclusive mode in the Breit frame and then boosted to the laboratory frame.
Thanks to the forward plug calorimeter installed for the
period 98-00, jets could be selected with a pseudorapity coverage in the laboratory frame of
$2< \eta_{\mathrm{jet}} < 3.5$. In order to enhance the contribution of possible BFKL effects, the
jets were required to have a $x_{\mathrm{jet}}> x$, where  $x_{\mathrm{jet}}$ is the ratio of the
longitudinal momentum of the jet and the proton momentum, and $x$ is the Bjorken variable. This requirement maximises
the phase space for BFKL evolution. In addition
the jets were required to have a transverse momentum squared 
$(E_T^{\mathrm{jet}})^2$ approximately equal to $Q^2$
in order to suppress DGLAP evolution as this cut leaves no room for evolution in $Q^2$. The events were selected in
the kinematic region $20<Q^2<100~\mathrm{GeV}^2$ and $0.0004<x<0.005$. The differential cross sections as a function of
$Q^2$, $x$, $E_T^{\mathrm{jet}}$ and $\eta_{\mathrm{jet}}$ are shown in figure~\ref{fjets}, compared
to NLO QCD calculations based on the DGLAP evolution, with the program DISENT. The data are
slightly above the theoretical NLO calculations, especially in the lowest $x$ bin, however the uncertainties
on the theory are still very large, as shown by the hatched bands. 
The variation of the calculations 
due to the change of renormalization scale is particularly large, which indicates the need for higher-oder calculations.  

\subsection{Very high-$x$ at HERA}

While HERA has provided precision results on structure functions at low-$x$, the high-$x$ region
has still to be explored, because of both limited statistics and
difficulty in reconstructing the kinematic variables. ZEUS has developed
a new method to select events at very high-$x$ and in the middle $Q^2$
region ($Q^2> 576~\mathrm{GeV}^2$). These events are characterized by a well
measured scattered electron or positron in the central part of the
calorimeter; and by a jet, very forward, close to the proton beampipe. 
As $x$ increases, the jet is more and more boosted in the forward direction
and eventually disappears in the beam pipe; the value of $x$ at which this occurs is $Q^2$ dependent. The kinematic of the event is reconstructed in this way. First the $Q^2$ of the event is determined from the electron variables
$E_e$ and $\theta_e$, which are
measured with good resolution. The events are then separated in those
with exactly one good reconstructed jet and those without any jet. For the 1-jet events, the
jet information is used to calculate the Bjorken $x$ variable from $E_\mathrm{jet}$
and $\theta_\mathrm{jet}$ and the double differential cross section
$d^2 \sigma/dx dQ^2$ is calculated in each $x,Q^2$ bin.
For the $0$-jet sample, it is assumed that the events come from very high-$x$, with a lower value $x_\mathrm{edge}$ which can be evaluated for each $Q^2$ bin
based on kinematic constraints. In this case the events are collected in a bin $x_\mathrm{edge} < x< 1$ and an integrated cross section 
$\int_{x_\mathrm{edge}}^1 (d^2 \sigma/ dx dQ^2) dx$ is calculated.

\begin{figure}
{\includegraphics[height=10cm]{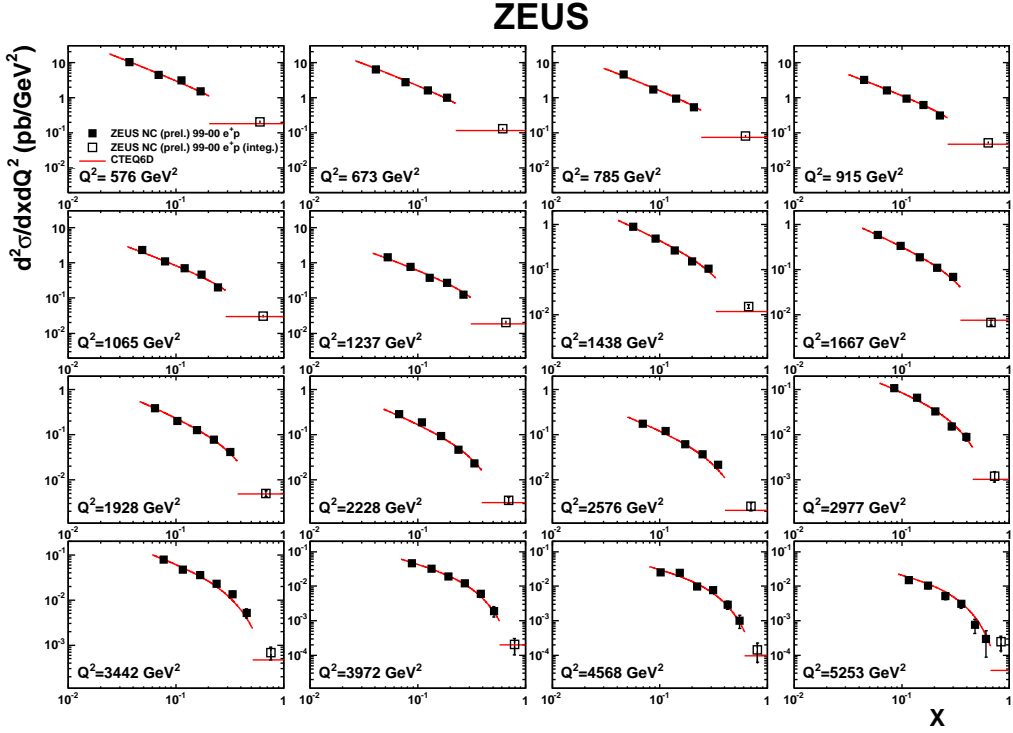}}
  \caption{Double differential cross sections (solid squares) for the 99-00 $e^+p$ data, as a function
of $x$ in $Q^2$ bins, compared to the CTEQ6D parton distributions. 
The last bin (open symbol) shows the integrated cross section over
$x$ divided by the bin width, i.e. $1/(1-x_{edge}) \cdot \int_{x_{edge}}^1 (d^2 \sigma/ dx dQ^2) dx$; the symbol is shown at the centre of the bin. 
In this bin the prediction is drawn as an horizontal
line. The error bars represent the quadratic sum of the statistical and
systematic uncertainties. }
  \label{highx}
\end{figure}

The result is shown in figure~\ref{highx}, where, for each $Q^2$,  the open squares in the last bin are
the integrated cross sections up to $x=1$ and the
closed squares show the double differential cross section in $x,Q^2$. The
precision in the last bin is comparable to the other bins.
For most of these highest $x$-bins, where there is no previous measurement,
the data tend to lie above the expectations from CTEQ6D~\cite{cteq}.
Precision measurements at high-$x$ at HERA will allow to constrain the valence
parton distributions, where up to now only fixed target experiments at low
$Q^2$ have provided experimental data.

\section{Diffraction}

\begin{figure}
{\includegraphics[height=7.5cm]{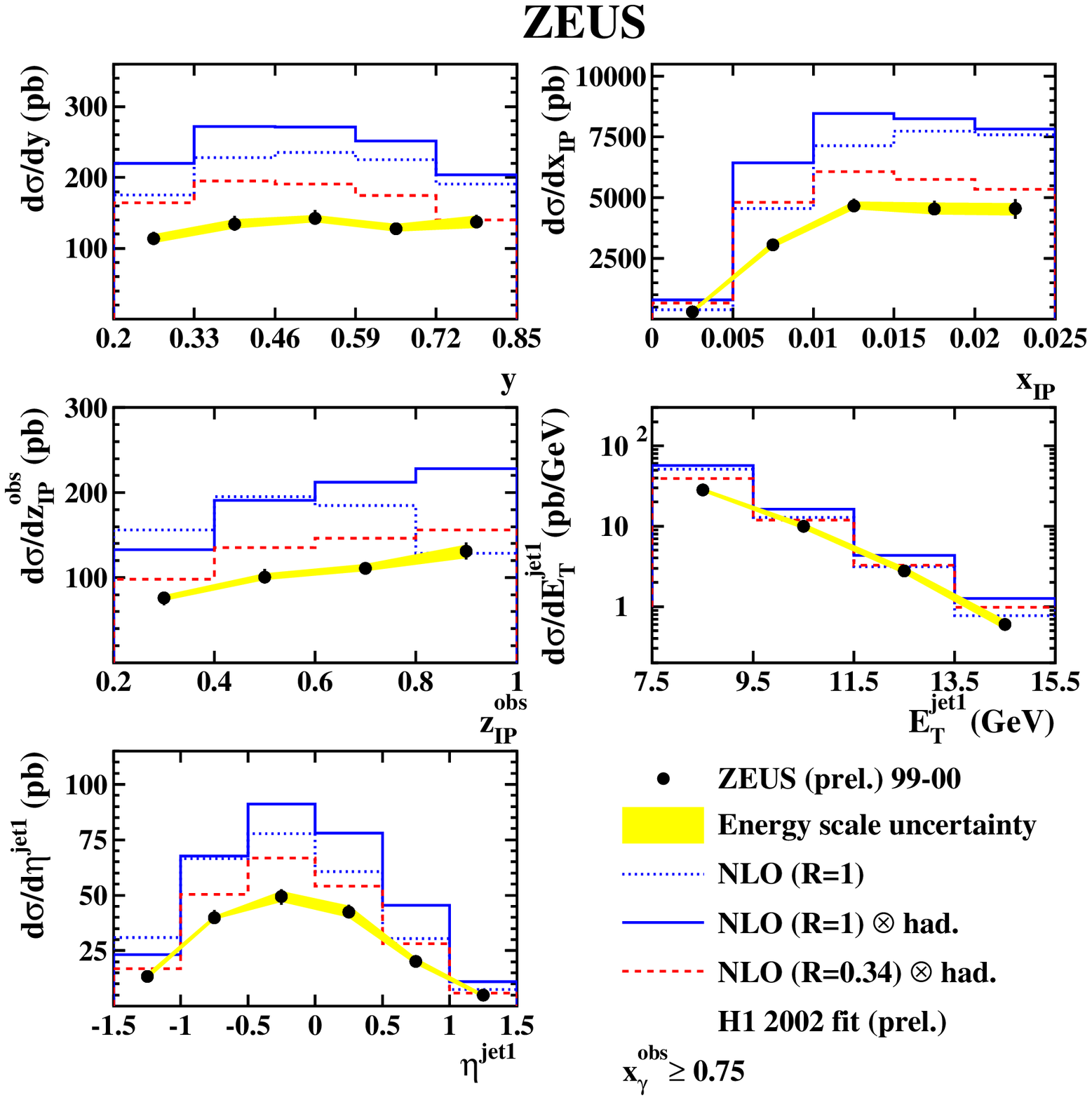}}
{\includegraphics[height=7.5cm]{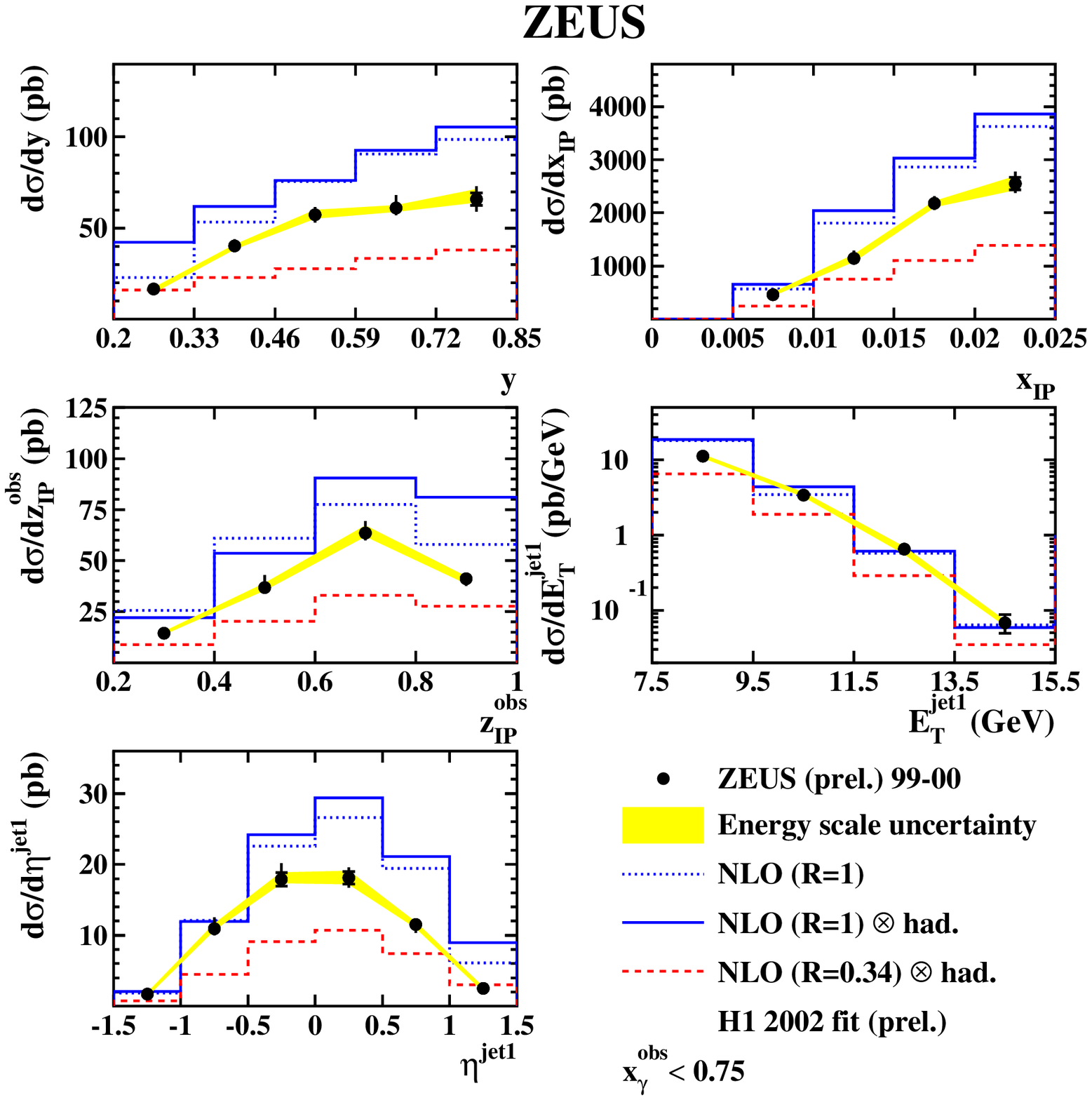}}
  \caption{Differential cross sections for diffractive dijet photoproduction
for a direct-enriched (left) and a resolved-enriched (right) sample. The variable
$\zpom$ is the longitudinal 
fractional momentum taken from the diffractive exchange by the dijet system.}
\label{phdiffdijets}
\end{figure}

At HERA, NLO QCD fits have also been performed to inclusive diffractive data
and  diffractive parton distribution functions (dPDFs) have been extracted. 
Assuming the QCD factorization theorem to be valid, in hard diffractive
processes the cross section can be factorized into the partonic
cross sections and these dPDFs, which are assumed to be universal. 
The dPDFs are  then used
to make predictions for various diffractive final states. Here I will
concentrate on diffractive photoproduction of dijets, which were measured
in the kinematic range $0.2 < y < 0.85$ and $\xpom<0.025$.
Jets were selected with
the longitudinally invariant $k_T$ algorithm and with
the asymmetric cuts $E_T>7.5,6.5$ GeV for the two jets. 
Double differential cross sections were measured separately in the 
region $x_{\gamma}^{obs}<0.75$, which is the
resolved-photon enriched region, and the region  $x_{\gamma}^{obs}>0.75$, which is
the region where direct-photon interactions dominate. The measured distributions in $y,\xpom,\zpom$ and $E_T$ and $\eta$ of the highest-$E_t$ jet are shown
at the hadron level in figure~\ref{phdiffdijets}. The data are compared to NLO calculations~\cite{klasenkramer} with
diffractive PDFs extracted from the H1 data~\cite{h12002}. 
In general the NLO calculations (shown by the line $\mathrm{R}=1$) reproduce 
the shape of the distributions, but their normalization is too high.

A reduction factor $\mathrm{R}=0.34$ has been calculated~\cite{kaidalov} in $p \bar p$ 
diffractive interactions, originating from interactions between 
spectator partons in the two hadronic beams. Such rescattering processes create additional particles that fill the large rapidity gap characteristic of diffractive
processes, causing a suppression of the measured cross sections and a break
of factorization.
A similar reduction could be
expected for resolved processes in which the photon behaves like a hadron. 
The dashed curve in figure~\ref{phdiffdijets}, corresponding to NLO predictions with $\mathrm{R}= 0.34$,
 is however too low compared
to the measured resolved-photon cross sections and  a suppression factor 
of $\mathrm{R} \simeq 0.5$ seems more appropriate.
A similar suppression factor seems to be needed also for direct
processes, as shown by the left plot of figure~\ref{phdiffdijets}, although
in this case the photon is more point-like and factorization is expected to
hold.

\begin{figure}
{\includegraphics[height=9cm]{lambda1520.epsi}}
{\includegraphics[height=9cm]{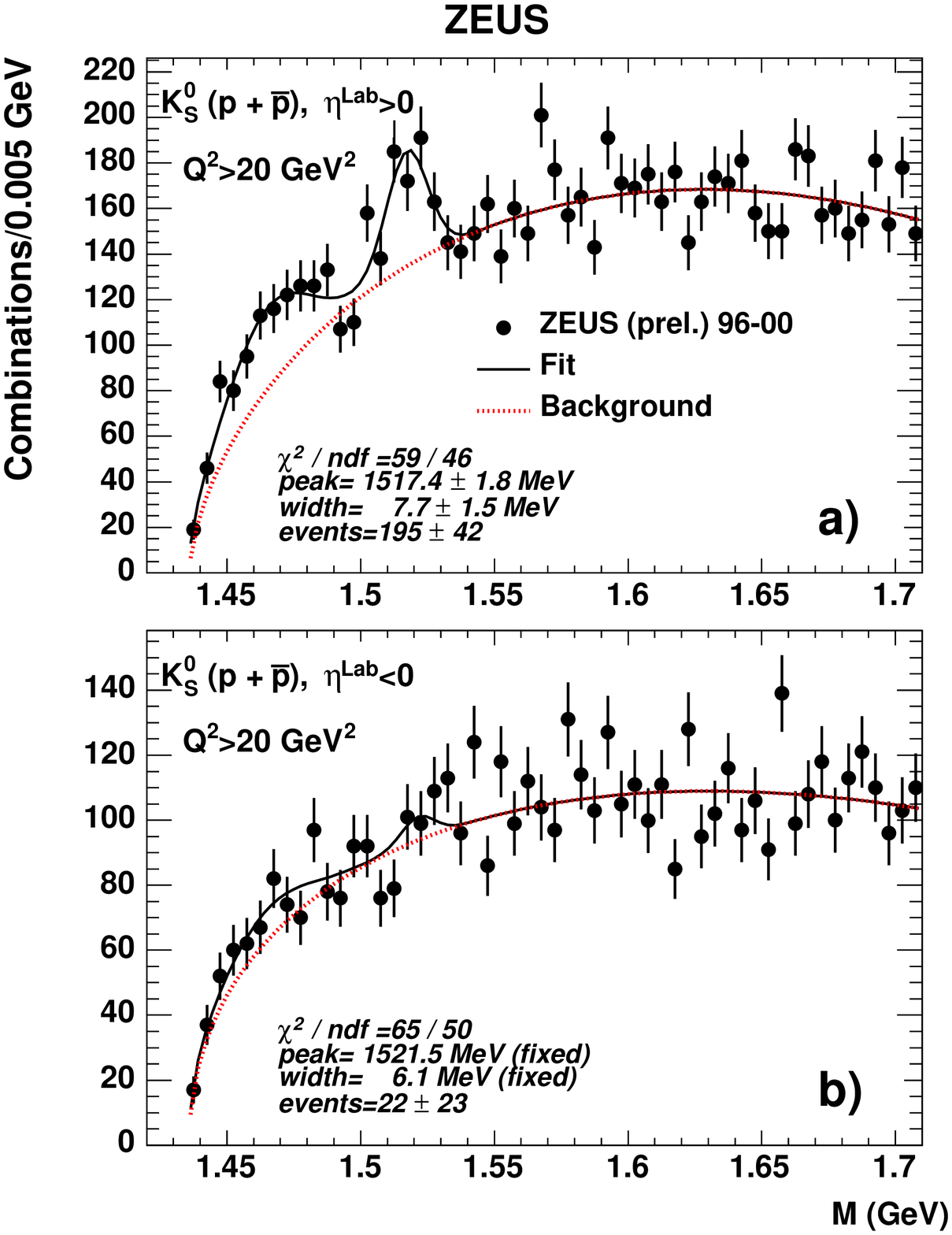}}
  \caption{Invariant mass spectra for the $\Lambda(1520)$ (left) and for the $\Theta^+$ (right),
for $Q^2>20~\mathrm{GeV}^2$ and divided in forward pseudorapidity region (top) and rear region (bottom). The
forward region corresponds to the direction of the proton.}
  \label{pentaquarks}
\end{figure}

\section{Exotic Final States}

Recently there has been many experiments reporting evidence of new
baryonic states consisting of five quarks. A number of experiments,
including ZEUS~\cite{pqstrange1}, have observed a narrow resonance decaying either to $nK^+$
 or $pK^0_S$ and with a mass around 1530 MeV, which could correspond to
the pentaquark state $\Theta^+=uudd \bar s$. Other experiments have reported negative 
searches for this state and ZEUS is at the moment the only high energy experiment
to have observed the $\Theta^+$ candidate. Recent studies from ZEUS have focused in trying to understand the production mechanism of this state. 

The $K^0_S p(\bar p)$ spectrum was studied for $Q^2>20~\mathrm{GeV}^2$ separately in the forward proton region, selecting
events with $\eta>0$, and in the rear region, $\eta<0$. The resulting spectra are shown in 
figure~\ref{pentaquarks}: the fitted number of events under the peak is higher in the region
closer to the proton remnant, compared to the rear region. This is not what is observed
in the production of the $\Lambda(1520) \rightarrow K^+p$, as shown in the same figure. This suggests
that the production mechanism for the $\Theta^+$ could be different from pure fragmentation
in the central rapidity region and could be related to
proton-remnant fragmentation.

ZEUS has also performed a search for two other exotic states: the $\theta_C= uudd \bar c$ observed
by H1~\cite{h1pqcharm} in the HERA I data: and the states $\Xi_{3/2}^{--}$ or  $\Xi_{3/2}^{0}$ with a mass around 1860 MeV decaying 
in $\Xi \pi$,  observed by the NA49 Collaboration~\cite{na49}. The $\Theta_c$ was
observed by H1 as a narrow resonance at a mass of $3099~\mathrm{GeV}$ 
decaying to $D^* p$, both in a $Q^2>1~\mathrm{GeV}^2$
sample and in a photoproduction sample: approximately $1\%$ of the selected $D^\ast$ mesons were observed to 
come from the decay of a $\Theta_C$. 
A similar analysis was performed by ZEUS~\cite{pqcharm} in an inclusive sample
with approximately 60,000 $D^\ast$ candidates, but no evidence for this state was found. 
Also a search~\cite{pqstrange2}
for the two states observed by NA49,
 decaying respectively to $\Xi^- \pi^-$ or $\Xi^- \pi^+$, was
performed and no resonance was found in a large mass range nor around the region of
1860 GeV. 

\begin{figure}
{\includegraphics[height=8cm]{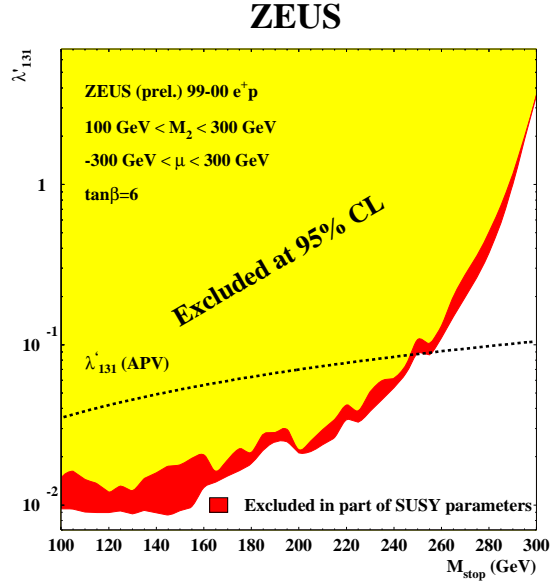}}
  \caption{Limits on the $R$-parity-violating coupling $\lambda^\prime_{131}$
as a function of the stop mass. The light region shows the excluded region 
for all considered SUSY scenarios, the dark area for a part of them. The 
dashed line is the limit from atomic-parity violation measurements.}
  \label{stoplimits}
\end{figure}

\section{$R$-parity-violating SUSY Searches}

HERA is the ideal place to look for possible exotic states originating
from the fusion of the initial positron/electron and a valence quark
in the proton. Among these, the stop can be produced 
from $e^+ d \rightarrow \tilde t$ via the $R$-parity-violating coupling 
$\lambda^\prime_{131}$. The stop can then decay as $\tilde t \rightarrow e^+ d$
or through the decay $\tilde t \rightarrow b \chi^+$,
giving a rich topology of final states. ZEUS has performed a search with
the 99-00 $e^+p$ data looking for final state topologies with one
positron and one jet (e-J) or multi jets (e-MJ) and one neutrino and multi
 jets ($\nu$-MJ). The branching ratios to the various channels depend on
the SUSY parameters, therefore a scan over a wide range of SUSY parameters
was performed. Selection cuts were designed to optimize  the 
signal sensitivity with respect to the main background, which comes from NC and 
CC interactions at high $Q^2$. No resonance was observed in the invariant mass of the
final state particles in the different topologies, therefore $95\%$ confidence 
level limits were
set on stop production. The results are shown in figure~\ref{stoplimits}
for the stop mass versus the coupling $\lambda^\prime_{131}$. The yellow (light) area
is the excluded region; the effect of changing the SUSY parameters in a
wide range and for $\tan \beta=6$,
as shown in the legend of the figure, is shown by the dark shaded band. The limits for masses 
up to 250 GeV improve on the low-energy limits
from atomic parity violation (APV) measurements and depend weekly on the
SUSY parameters.

\begin{figure}
{\includegraphics[height=8cm]{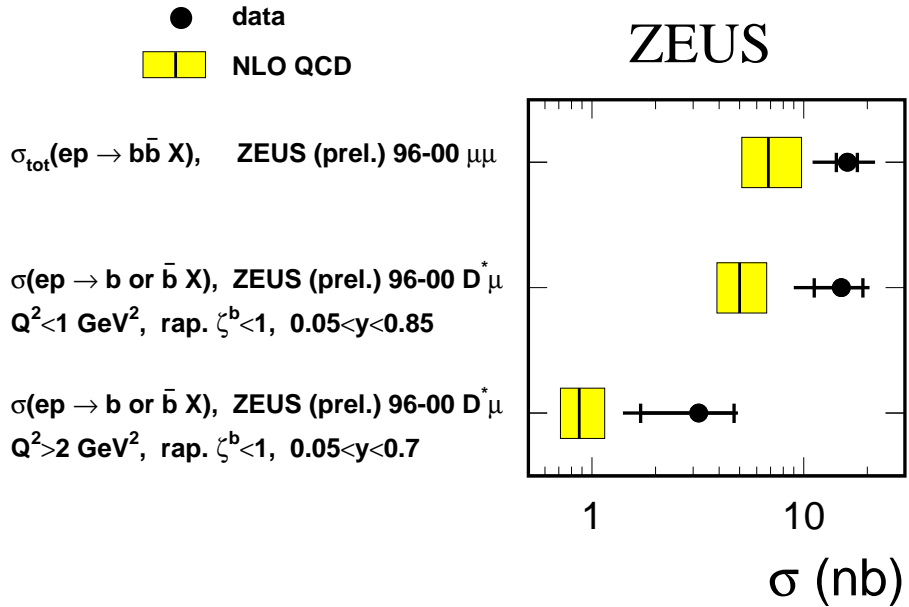}}
  \caption{Beauty cross section from dimuon and $D^\ast+\mu$ tagged events.
The measurements (dots) are compared to the NLO QCD calculations.}
  \label{beautyxsec}
\end{figure}

\section{Heavy Flavours}

Beauty production has been measured by ZEUS both in photoproduction
and in DIS. The hard scale provided by the mass of the $b$-quark allows
to compare to NLO QCD calculations.
A new measurement based on $121~\mathrm{pb}^{-1}$ of HERA I data in which two muons
are observed in the final state was presented at this workshop. Tagging both
muons coming from the semileptonic $b$ decays allows a better suppression of the
backgrounds from charm and light flavour production. This also means that 
muons, and therefore $b$'s, can be selected at low $p_T$ and in a large $\eta$ range, which implies less extrapolation in deriving the total cross section.
In addition the normalizations of the various backgrounds can be constrained dividing the sample
into high- and low-mass, isolated and non-isolated, like- and unlike-sign muon pairs. 

From the visible cross section,  
a total cross section for beauty production at HERA is extracted and compared to NLO QCD predictions from the sum of the two calculations FMNR (for photoproduction) and HVQDIS (for DIS), obtaining:
\begin{eqnarray}
\sigma^b_{tot}(ep \rightarrow b \bar b X)[318~\mathrm{GeV}] & = &
16.1 \pm 1.8 (stat.)^{+5.3}_{-4.8} (syst.)~\mathrm{nb}, \\
\sigma^b_{tot}(\mathrm{NLO}) & = & 
6.9^{+3.0}_{-1.8}~\mathrm{nb},
\end{eqnarray}
where the uncertainties on the theory were evaluated changing the 
renormalization and factorisation scales by a factor 2 and varying the $b$
mass between 4.5 and 5 GeV.

The result is shown in figure~\ref{beautyxsec}, where also
previous results from ZEUS, based on  $D^\ast + \mu$ samples~\cite{dstarmu}, are shown.
These samples also allow measurement of the cross section to be made
for low transverse momentum of the $b$'s. From the figure it looks like
there is the tendency for these cross sections to be above 
NLO QCD predictions,
while for higher $Q^2$ and higher $p_T$ there is good agreement between
theory and measurements (see e.g. ~\cite{h1bpaper}).

\begin{figure}[h]
{\includegraphics[height=9cm]{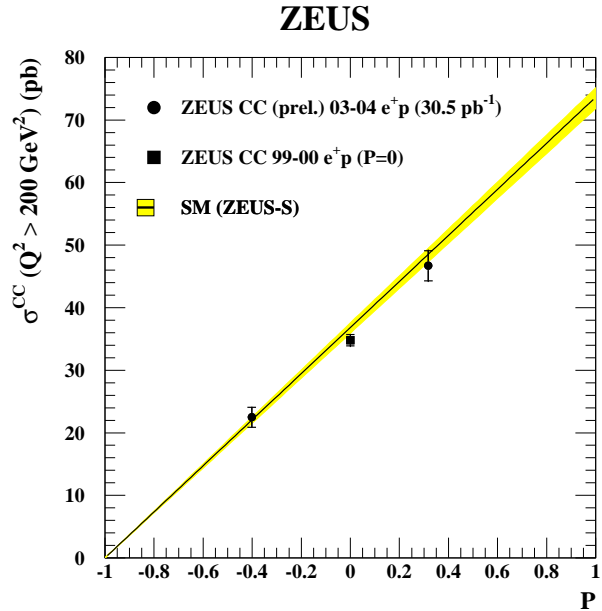}}
  \caption{Charged current total cross sections as a function of
the longitudinal polarization of the positron beam, 
compared to the SM predictions with the ZEUS-S PDFs. The square symbol
shows the HERA I results from the published unpolarized $e^+p$ data.}
  \label{ccxsec}
\end{figure}

\section{HERA II}

\subsection{Cross sections at high $Q^2$}

The HERA II running phase has started, $e^+p$ collisions with
longitudinally polarized positrons have been collected in 
2003-2004 and in 2004-2005
the machine has switched to polarized electron beams.
Cross sections at high $Q^2$ were measured with the first 2003-2004 data
with polarized positron beams~\cite{ccpol}, with a mean luminosity weighted
polarization $P=-40.2\%$ for $16.4~\mathrm{pb}^{-1}$ of collected data, and 
$P=+31.8\%$ 
for $14.1~\mathrm{pb}^{-1}$ of integrated luminosity. The total charged 
current cross section is predicted in the Standard Model to have a linear 
dependence on the polarization, becoming zero for left-handed positrons
(or right-handed electrons).

The measured CC cross sections for the two polarization values
are shown in figure~\ref{ccxsec}: the linear dependence on $P$ is
clearly visible and the points are in good agreement with the SM prediction,
which is calculated with the ZEUS-S parton densities for the proton. 
The effect of the longitudinal polarization in the NC cross sections 
can only be seen at very high $Q^2$ with a high integrated luminosity. 
A marginal effect could anyway already be seen in 
these data~\cite{ccpol}.

\subsection{Charm cross sections}

\begin{figure}[h]
{\includegraphics[height=8cm]{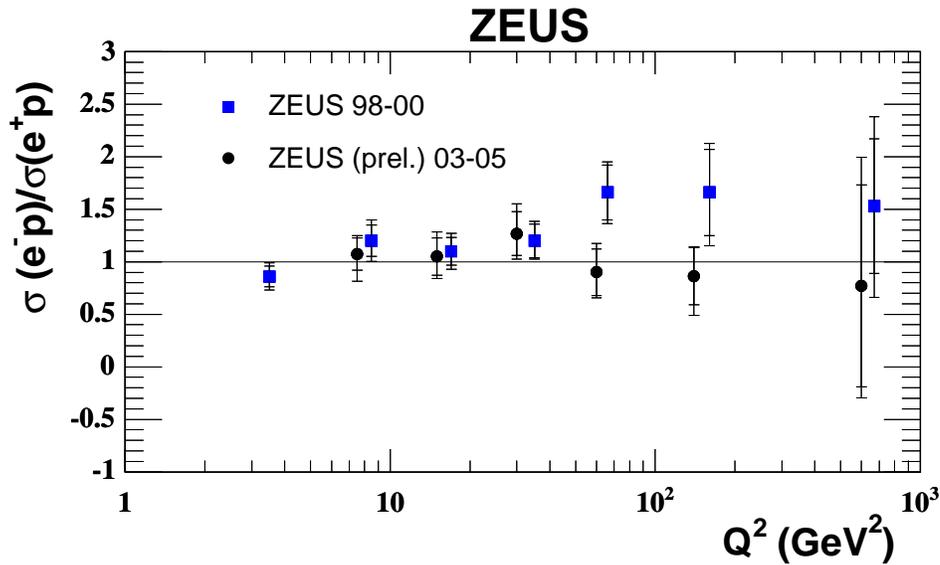}}
  \caption{The ratio of the cross sections for $D^\ast$ production
in $e^-p$ to the one in $e^+p$,
as measured in the 03-05 data and in the 98-00 data. The inner error bars show the statistical uncertainty, the outer error bar the systematic uncertainty.
Many systematic effects cancel in the ratio.}
  \label{dstarratio}
\end{figure}

Charm is produced in DIS mainly through the boson-gluon-fusion mechanism 
and is therefore directly sensitive to the gluon density in the proton.
The cross section for charm was measured in $40~\mathrm{pb}^{-1}$ of $e^+p$ and 
$33~\mathrm{pb}^{-1}$ of $e^+p$ data collected in 2003-2005, as a function of
$Q^2$ in the range $5< Q^2 < 1000~\mathrm{GeV}^2$. Charm was tagged using the 
$D^\ast \rightarrow D^0 \pi_S \rightarrow K \pi \pi_S$ decay, where
tracks were reconstructed with the central tracking detector  and
the newly installed silicon microvertex detector. Approximately 1200
$D^\ast$ candidates were selected in each sample, $e^+p$ and $e^- p$.
The ratio of the cross sections for the two samples was measured, as shown in 
figure~\ref{dstarratio}, as a function of $Q^2$. This ratio is particularly
interesting as, with the previous published HERA I data~\cite{charmhera1}
(based on $17~\mathrm{pb}^{-1}$ of $e^- p$ and $65~\mathrm{pb}^{-1}$ of $e^+p$),
the $D^\ast$ production rate was observed to be higher in $e^- p$ than
in $e^+ p$, as shown also in the figure. This ratio was found to be
$1.67 \pm 0.21$ (only statistical error) in the range $40 < Q^2 < 1000~\mathrm{GeV}^2$
and such phenomenon is not expected from any
physics process. With the higher statistics in the $e^-p$ data, 
collected up to now at HERA II, 
this effect is not confirmed and the
ratio agrees with unity through the whole range of $Q^2$ measured.

\section{Conclusion}
The ZEUS Collaboration is completing the analysis of the HERA I
data and starting to look at the HERA II data. Recent highlights
have been described here; many more results and details can be found
in these proceedings from the parallel sessions contributions.

\begin{theacknowledgments}
I would like to thank my ZEUS colleagues, and in particular Matthew Wing
and Rik Yoshida, for the preparation of this talk; and Prof. Wesley H. Smith 
for the excellent organization of this Workshop.
\end{theacknowledgments}



\begin{thebibliography}{15}

\bibitem[Coll.(2005)]{zeusjets}
ZEUS Collaboration, {DESY-05-071} (2005).

\bibitem[Glasman(2005)]{claudia}
C.~Glasman, {these proceedings and references therein} (2005).

\bibitem[Mueller(1993)]{mueller}
A.~H. Mueller, \emph{J. Phys. G}, \textbf{19}, 1463--1468 (1993).

\bibitem[J.~Pumplin(2002)]{cteq}
J.~Pumplin et al., \emph{JHEP}, \textbf{0207}, 012 (2002).

\bibitem[Klasen and Kramer(2004)]{klasenkramer}
M.~Klasen, and G.~Kramer, \emph{Eur. Phys. J}, \textbf{C38}, 93 (2004).


\bibitem[H1(2002)]{h12002}
H1 Collaboration, {paper 980 submitted to ICHEP02} (2002).


\bibitem[Kaidalov(2003)]{kaidalov}
A.~B. Kaidalov, \emph{Phys. Lett.}, \textbf{B567}, 61 (2003).

\bibitem[ZEUS(2004{\natexlab{a}})]{pqstrange1}
ZEUS Collaboration, \emph{Phys. Lett.}, \textbf{B591}, 7--22 (2004{\natexlab{a}}).

\bibitem[H1(2004)]{h1pqcharm}
H1 Collaboration, \emph{Phys. Lett.}, \textbf{B588}, 17--28 (2004).

\bibitem[NA49(2004)]{na49}
NA49 Collaboration, \emph{Phys. Rev. Lett.}, \textbf{92}, 042003 (2004).

\bibitem[ZEUS(2004{\natexlab{b}})]{pqcharm}
ZEUS Collaboration, \emph{Eur. Phys. J.}, \textbf{C38}, 29--41 (2004{\natexlab{b}}).

\bibitem[ZEUS(2005)]{pqstrange2}
ZEUS Colaboration, {DESY-05-018} (2005).

\bibitem[ZEUS(2004{\natexlab{c}})]{dstarmu}
ZEUS Collaboration, {paper 5-0342 submitted to ICHEP04} (2004{\natexlab{c}}).

\bibitem[H1(2005)]{h1bpaper}
H1 Collaboration, \emph{Eur. Phys. J.}, \textbf{C41}, 453 (2005).

\bibitem[ZEUS(2004{\natexlab{d}})]{ccpol}
ZEUS Collaboration, {paper 4-0256 submitted to ICHEP04} (2004{\natexlab{d}}).

\bibitem[ZEUS(2004{\natexlab{e}})]{charmhera1}
ZEUS Collaboration, \emph{Phys. Rev.}, \textbf{D69}, 012004 (2004{\natexlab{e}}).

\end{thebibliography}

\end{document}